\documentclass{epl}

\usepackage{graphicx,amsmath,amssymb}
\newcommand{\bb}{\begin{equation}}
\newcommand{\en}{\end{equation}}

\def\vf{{\bf f}}
\def\vv{{\bf v}}
\def\rp{{\bf r_\perp}}
\def\vE{{\bf E}}
\def\vq{{\bf q}}
\def\sp{{\sigma^+}}
\def\sm{{\sigma^-}}
\def\kp{{\kappa^+}}
\def\km{{\kappa^-}}
\def\kpm{{\kappa^\pm}}
\def\spm{{\sigma^\pm}}

\title{Fluctuations of a driven membrane in an electrolyte}
\author{D. Lacoste\inst{1}, M. Cosentino Lagomarsino\inst{2,3},
 and JF. Joanny\inst{2}}

\institute{
 \inst{1} Laboratoire Physico-Chimie Th\'eorique, ESPCI, 10 rue Vauquelin 75005 Paris, France  \\
 \inst{2}  UMR 168 / Institut Curie, 26 rue d'Ulm 75005 Paris, France \\
 \inst{3} Universit\`a degli Studi di Milano, Dip.
    Fisica, Via Celoria 16, 20133 Milano, Italy
}


\begin{document}

\maketitle

\begin{abstract}
  We develop a model for a driven cell- or artificial membrane in an
  electrolyte. The system is kept far from equilibrium by the application of a
  DC electric field or by concentration gradients, which causes ions to flow
   through specific ion-conducting units (representing pumps,
    channels or natural pores). We consider the case of planar
  geometry and Debye-H\"{u}ckel regime, and obtain the membrane equation of motion
   within Stokes hydrodynamics.
  At steady state, the applied field causes an accumulation of charges close
  to the membrane, which, similarly to the equilibrium case, can be described
  with renormalized membrane tension and bending modulus. However, as opposed
  to the equilibrium situation, we find new terms in the membrane equation of
  motion, which arise specifically in the out-of-equilibrium case. We show
  that these terms lead in certain conditions to instabilities.
\end{abstract}

\section{Introduction}
Phospholipid membranes are a major constituent of the cell.  They
cover the surface of all organelles and play an important role in
many fundamental cellular processes such as intracellular
transport~\cite{Edidin03}. At a mesoscopic scale, the equilibrium
properties of fluid membranes can be accounted for by the
so-called Helfrich Hamiltonian, which describes the mechanics of
the membrane in terms of parameters, such as membrane tension,
bending modulus and spontaneous curvature. These mechanical
parameters strongly depend on the electrostatic charge of the
phospholipid molecules. The electrostatic contributions to the
bending modulus and surface tension for a fluid membrane have been
obtained by several authors \cite{helfrich,duplantier,pincus}
using free energy calculations for simple geometries (sphere,
cylinders, planes).

\begin{figure}
{\par \rotatebox{0}{\includegraphics[scale=0.5]{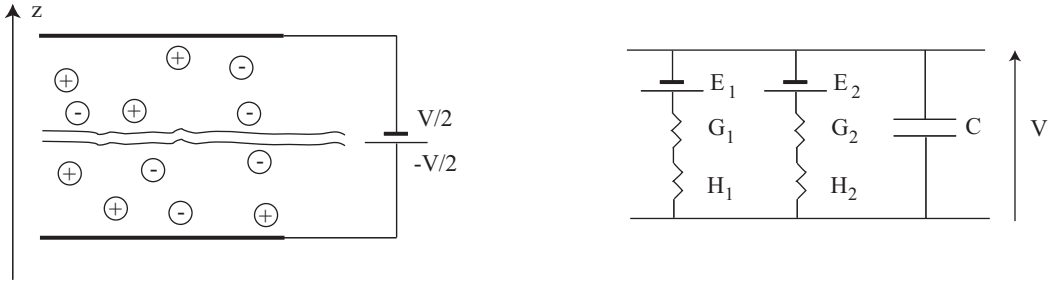}}
\par}
\caption{Left: Sketch of a quasi-planar fluid membrane embedded in
an electrolyte in the presence of an applied potential difference
and concentration gradients (not represented). Right: equivalent
electrical circuit, with the contribution of the two ions $k=1,2$
in parallel, and for each ion, two conductances ($G_k$ for the
membrane and $H_k$ for the electrolyte) and an electromotive force
$E_k$ in series. When the finite thickness of the bilayer is
included, there is also a capacitor $C$ in parallel which contains
contributions from Debye layers and from the membrane.}
\label{fig:sketch}
\end{figure}

Real membranes are non-equilibrium systems.  They are in general
active in the sense that they are constantly maintained out of
equilibrium either by active proteins (such as ATP-consuming
enzymes) inside the membrane or by an energy flow due to external
parameters (such as a lipid flux~\cite{girard}).  An example of
model active membrane has been proposed in
Ref.\cite{PRE_JBManneville}. In these experiments, a giant
unilamellar vesicle is activated by the inclusion of
bacteriorhodopsin pumps, which transfer protons unidirectionally
across the membrane by undergoing light-activated conformational
changes. In the same work, a hydrodynamic theory has also been
developed to calculate the nonequilibrium fluctuations of the
membrane, induced by the activity of the pumps.
In this paper, we study theoretically an electrically neutral
membrane containing passive ion channels in an electrolyte
solution and driven out of equilibrium by the application of a DC
electric field or by ion concentration gradients~\cite{leonetti}
(Fig.~\ref{fig:sketch}).
As in refs.~\cite{PRE_JBManneville,lacoste,lomholt}, we use a
generalized hydrodynamic description appropriate for low Reynolds
number, that we couple to the electrokinetic equations
\cite{armand} within Debye-H\"{u}ckel approximation. We calculate
the undulation fluctuations by expansion around the planar state
of the membrane. With minor modifications, our method can also be
used to calculate the electrical contribution to the fluctuations
of a membrane containing pumps. This contribution, was not taken
into account in ref.~\cite{PRE_JBManneville}. We first consider
the case of a membrane of zero thickness, and then generalize the
model to the case of a bilayer with a finite thickness and a
finite dielectric constant, lower than that of the solvent. The
later model is more realistic since it contains both capacitive
effects and ions transport \cite{hille,kandel}.

\section{Membrane of zero thickness driven by the application
of an electrostatic potential difference} \label{zero thickness}
In this section, we compute the electrostatic potential $\phi$,
and then use the electrostatic forces as a source term
in the hydrodynamic equations to obtain the membrane equation of
motion. We use a linear theory and assume a steady state. The
quasi-planar membrane is located in the plane $z=0$
(Fig.~\ref{fig:sketch}), it is embedded in an electrolyte and
carries passive channels for two types of monovalent ions. There
is an imposed potential difference $V$ across the system of length
$L$. The boundary conditions for the potential are therefore
$\phi(z= \pm L/2)= \pm V/2$ and $\partial_z \phi(z\to 0^\pm)=0$
due to the assumption of a vanishing membrane dielectric constant.
We denote $c_k$ the concentrations of the two ions, where the
index $k$ is 1 for the positive ion ($z_1=1$) and is 2 for the
negative ion ($z_2=-1$). Far from the membrane these
concentrations are fixed, so that $c_k^{\pm}(z= \pm L/2)=n^{\pm}$,
where the superscript $+$ (resp. $-$) denotes $z>0$ (resp. $z<0$).
Everywhere in the electrolyte we  linearize the concentrations around
these bulk values. A point on the membrane is
characterized by a 2D vector field $\rp$ and for small undulations
a height function $h(\rp)$. The inverse Debye-H\"{u}ckel length is
$\kp$ on the positive side (resp. $\km$ on the negative side) with
$\kappa^{\pm 2}=2 e^2 n^\pm / \epsilon k_B T$. Within a linear
theory for the concentrations and the potential, the fluxes of the
positive ions of bulk diffusion coefficient $D_1$ and of the
negative ions of diffusion coefficient $D_2$ are proportional to
the ion chemical potential gradients
\begin{eqnarray}\label{ionfluxes}
{\bf J}_1^{\pm} &=& -n^{\pm} D_1 \left[ \nabla \rho_1^\pm +
\nabla \psi^\pm \right] \\
\nonumber {\bf J}_2^{\pm} &=& -n^{\pm} D_2 \left[ \nabla \rho_2^\pm
- \nabla \psi^\pm \right],
\end{eqnarray}
where $\rho_k$ are normalized concentrations such that
$c_k^\pm=n^\pm (1 + \rho_k^\pm)$, and $\psi=e \phi / k_B T$ is the
normalized potential. The charge density is $\rho^\pm=e n^\pm
(\rho_1^\pm - \rho_2^\pm)$; it obeys Poisson's equation. The
conservation of both types of ions in an incompressible fluid
imposes that $\partial c_k /
\partial t =- \nabla \cdot {\bf J}_k$, and therefore at steady state
the ions fluxes are constant. At the membrane surface, the normal
bulk flux of each ion species is equal to the flux through the
membrane. Throughout this paper, we assume that the membrane has a
constant conductance per unit area $G_k$ for the two types of
ions. In general, the conductance depends in a non linear way on
the electrostatic potential difference across the
membrane~\cite{hille,kandel}, but here as a first step we assume
linear conductivity. In the case of symmetric concentrations
$n^+=n^-$, the charge density is an odd function of $z$ and the
electric field an even function of $z$, but no such symmetry is
present in the general case when $n^+ \neq n^-$.

The electrostatic potential and the charge density on the positive
(resp. negative) side depends on the surface charge on the
positive electrode (on the same side) $\sigma^+$ (resp.
$\sigma^-$), which is defined as $\sigma^\pm=\pm \epsilon
\partial_z \phi(z=\pm L/2)$. We find that
\begin{eqnarray}\label{surfacechargedef}
\sigma^+ & = & -\frac{1}{{\kp}^2}
\left( \frac{i_1}{D_1}+ \frac{i_2}{D_2} \right), \\
\nonumber \sigma^- &=& \frac{1}{{\km}^2} \left( \frac{i_1}{D_1}+
\frac{i_2}{D_2} \right),
\end{eqnarray}
where $i_1=e {\bf J}_1 \cdot {\bf {\hat e}_z}$ and $i_2=-e {\bf
J}_2 \cdot {\bf {\hat e}_z}$ are the electric currents carried by
ion species 1 and 2. Note that the electroneutrality is satisfied,
since $\int_0^\infty \rho^+(z) dz=-\sigma^+$ and $\int_{-\infty}^0
\rho^-(z) dz=-\sigma^-$. The non-vanishing charges of the
electrodes compensate the overall charge of the electrolyte. The
electric currents (for $L \kappa^\pm \gg 1$) are given by
\begin{equation}\label{current}
i_k= \frac{-G_k v_k }{1+G_k L k_B T/D_k n e^2},
\end{equation}
where $v_k=V-E_k$, $E_k=-k_B T \log (n^+ / n^-) / e z_k$ is the
Nernst potential of ion $k=1,2$ and $n= 2 n^+ n^- / (n^+ + n^-)$.
This equation is consistent with the usual electrical
representation of ion channels, shown in fig.~\ref{fig:sketch}
with the contribution of the two ions in parallel, and for each
ion, two conductances $G_k$ and $H_k$ and an electromotive force
$E_k$ in series~\cite{hille,kandel}. $H_k$ corresponds to the bulk
conductance per unit area of each ion $H_k=D_k n e^2 / k_B T L$,
which can be itself decomposed in two conductances in series for
each side of the electrolyte.

\begin{figure}[htbp]
  \centering
\rotatebox{-90}{\includegraphics[scale=0.5]{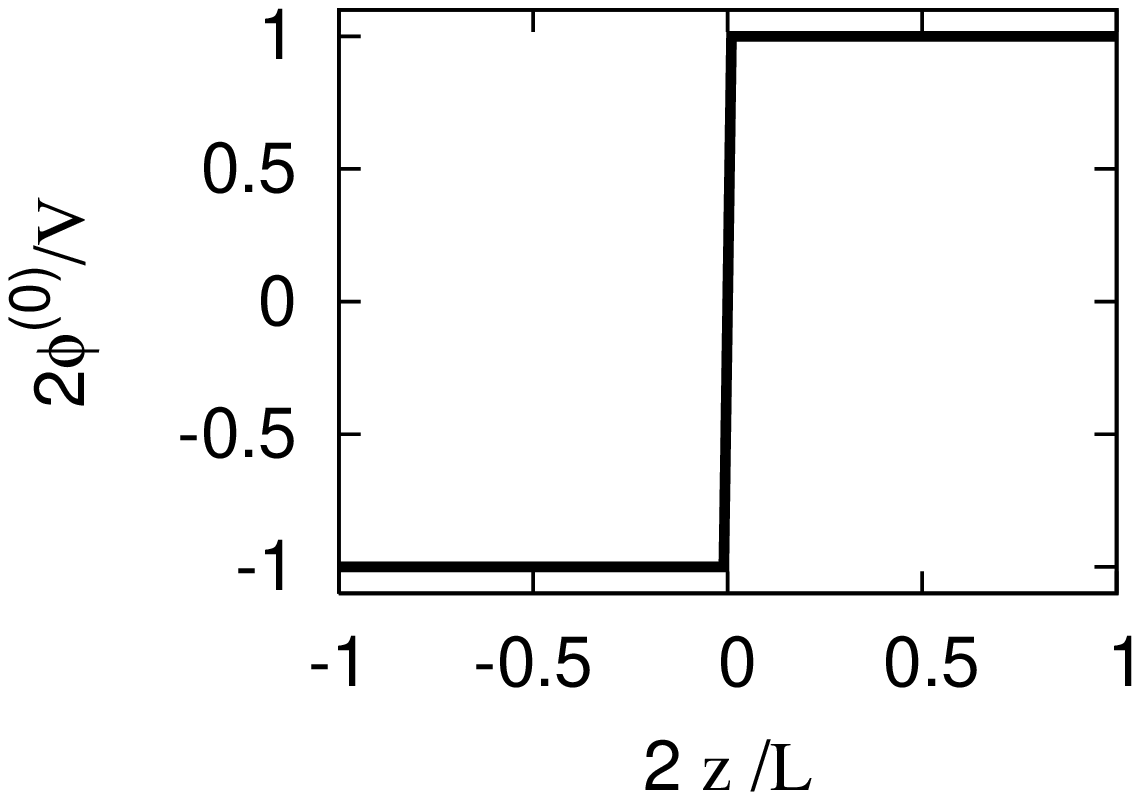}}
\rotatebox{-90}{\includegraphics[scale=0.5]{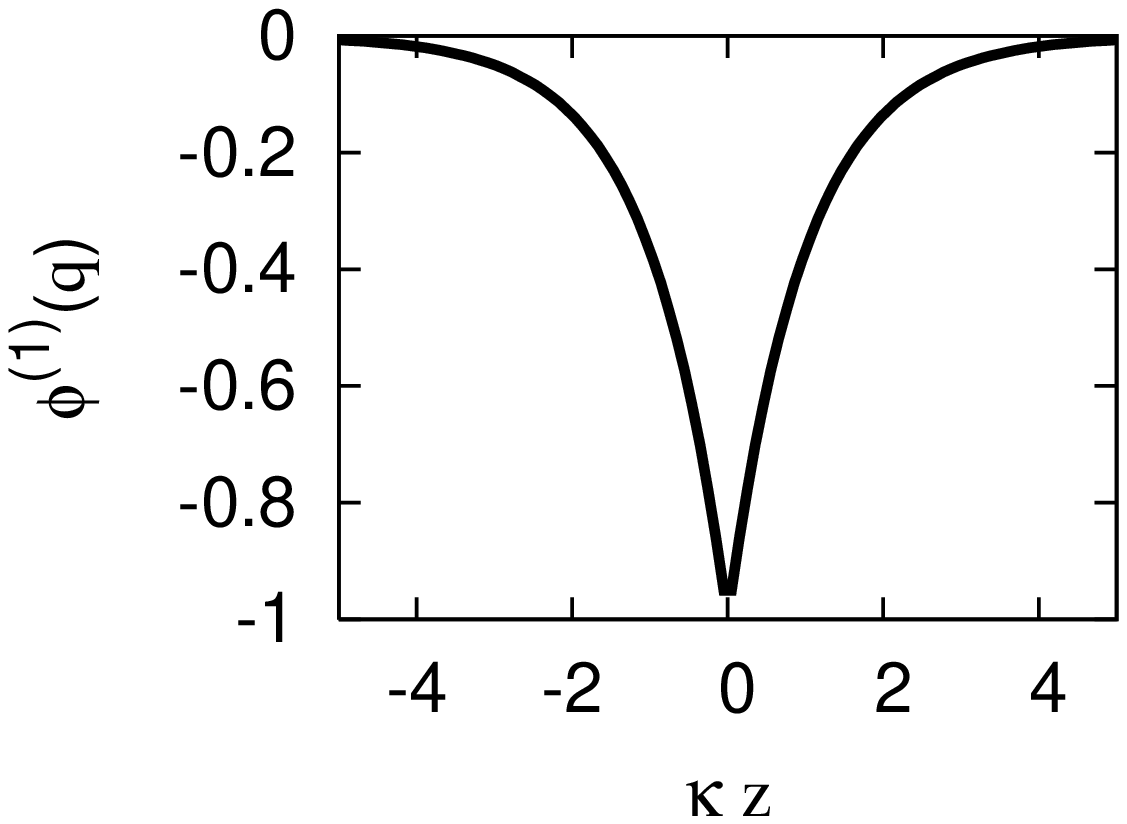}}
\rotatebox{0}{\includegraphics[scale=0.5]{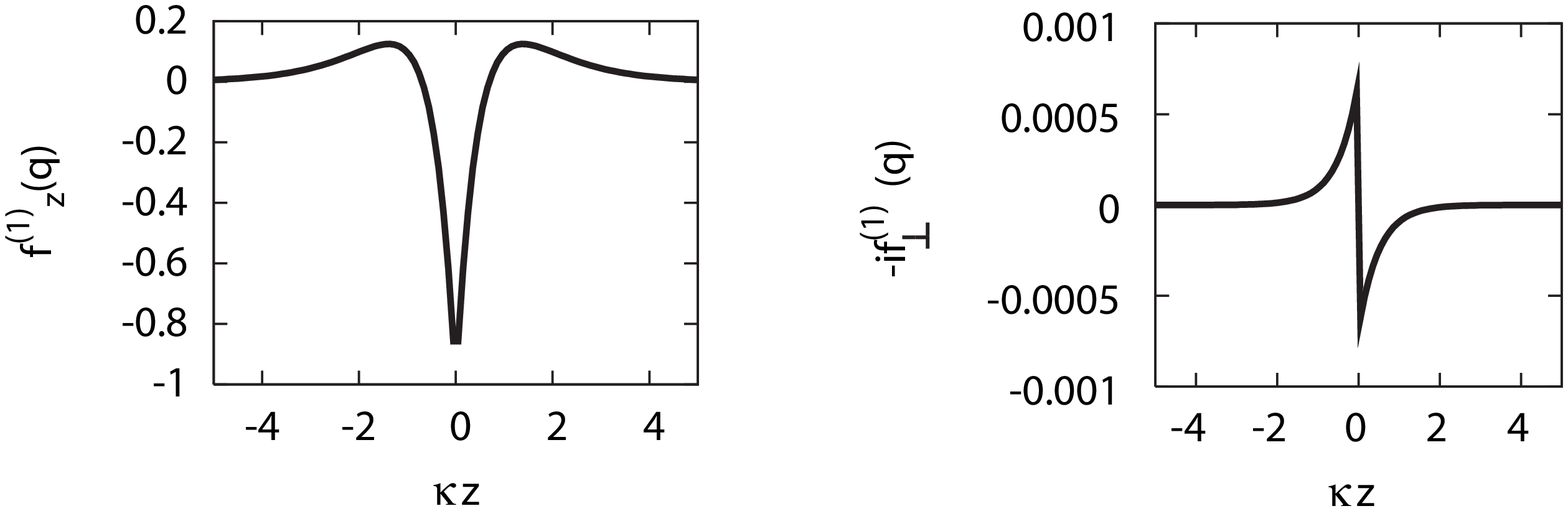}}
\caption{Solutions of the electrokinetic equations for a
    membrane of zero thickness and symmetric ions concentrations $n^-=n^+=n$.
    The parameters used in this figure are
    $V=50$mV, $L=1\mu$m, $G_1=G_2=10 \Omega^{-1}$/m$^2$, $D_1=D_2=10^{-5}$cm$^2/s$ and
    $n=16.6$mM. Top: Left: Normalized potential at zeroth order
    $2\phi^{(0)}(z)/V$ as function of $2z/L$.
    Right: Normalized potential at first order in membrane undulation
    $\phi^{(1)}(q,z)$ for a wavevector
    $q=10^6$m$^{-1}$, the normalization factor is $h(q) \sigma
    4 \pi/\epsilon$. Bottom: Left: Normalized force along $z$ at first order
    $f^{(1)}_z(q,z)$. Right: Normalized force along ${\bf q}$
    at first order
    $-i f^{(1)}_\perp(q,z)$ (where $i$ is the imaginary number).
    This is for the same wavevector and the normalization
    factor for the forces is  $h(q) \sigma^2 \kappa^2
    4 \pi/\epsilon$.}
 \label{fig:2}
\end{figure}

We now consider fluctuations to first order in the membrane
undulation $h$, and in a quasi-steady state, assuming that the
pulsation $\omega<< D_k \kappa^2$. The calculations are
conveniently carried out using Fourier transforms, defined as
$f(\vq,z) = \int_{-\infty}^{\infty} d^2 \rp e^{i\vq
  \cdot \rp} f(\rp,z)$, where $\vq$ is the transverse wave vector for membrane
fluctuations. Rotational symmetry imposes that all the fields in our
problem depend only on $q=|\vq|$. We find that, the first order
correction to the fluxes given by Eq.~\ref{ionfluxes} vanishes,
which indicates that there is no correction to the ion
conductivities due to membrane fluctuation at this order. More
precisely, the first-order steady-state solution of our
electrokinetic equations is then $\phi^{(1)}(q,z)=\pm \kappa^\pm
\sigma^\pm h(q) \exp(-\kappa_q^\pm z)/ \epsilon \kappa_q^\pm$ (see
Fig.~\ref{fig:2}) and $\rho^{(1)}(q,z)= \mp {\kappa^\pm}^3 h(q)
\exp(-\kappa_q^\pm z) \sigma^\pm / \kappa_q^\pm$, where
$\kappa_q^\pm =(q^2+{\kappa^\pm}^2)^{1/2}$.
%
%
%
Having determined perturbatively the ion concentrations and the
electric field, we study the membrane undulations imposed by the
hydrodynamics of the surrounding electrolyte. The velocity field
is obtained from the Stokes equation in the presence of a force
density $\vf$ acting on the fluid~\cite{seifert},
\begin{equation}\label{Stokes}
\eta \nabla^2 \vv - \nabla P + \vf=0 \ ,
\end{equation}
with the incompressibility condition $\nabla \cdot \vv=0$.  Away
from the membrane, the boundary condition is $\vv(\rp,z\to
\infty)=0$. We assume an absence of permeation through the
membrane and the no-slip condition $\vv(\rp,z=0)=\partial_{t}
h(\rp,t) \, \bf{\hat{e}}_z$.  Here, the active forces
\cite{lacoste,lomholt} only act on the ions that are transported
across the membrane, i.e. $\vf(\rp,z)=\rho(\rp,z) {\bf E}(\rp,z)$.
The zeroth order contribution to this force $\vf^{(0)}$ is
balanced by the pressure gradient of Eq.~\eqref{Stokes}. One can
then derive the fluid velocity $\vv$ at the membrane surface by
solving perturbatively Eq.~\eqref{Stokes}. For $q \ll \kpm$, and
up to first order in the membrane undulation, we find that
$v_z(q,z=0)=v_z^+(q,z=0)+v_z^-(q,z=0)$ with
\begin{equation}\label{velocities_ out of equil}
v_z^\pm (q,z=0)= \frac{q h(q)}{\epsilon \eta} \left( -\frac{3}{8}q
\frac{\spm^2}{\kpm} + \frac{1}{2} q^2 \frac{\spm^2}{\kpm^2} -
\frac{3}{32} q^4 \frac{\spm^2}{\kpm^3} \right) .
\end{equation}
This normal velocity could be derived from an effective free
energy density
\begin{equation}
\label{freeenergy} F_{eff}(q)=[\Sigma q^2 + \Gamma q^3 + K q^4]
h^2(q)/2
\end{equation}
with $\Sigma=3 (\sp^2/ \kp + \sm^2/\km) / 2 \epsilon$, $K=3 ( \sp^2/
\kp^3 + \sm^2/\km^3) / 8 \epsilon$ and $\Gamma= -2 (\sp^2/ \kp^2 +
\sm^2/\km^2) / \epsilon$. Thus, if nonthermal noise can be
neglected, the fluctuation spectrum is $\langle \vert h(q) \vert^2
\rangle =k_B T /\left((\Sigma_0+ \Sigma) q^2 +
 \Gamma q^3 +(K_0+K) q^4 \right)$,
where we have assumed that the membrane has, in addition to the
electrostatic contribution, a bare tension $\Sigma_0$ and a bare
bending modulus $K_0$.

Let us now discuss the physical significance of this result:

\noindent (i) The first important point is that this derivation of
the fluctuation spectrum is independent of the origin of the ion
fluxes. In particular, it applies also to the case where the ion
fluxes are produced {\it internally} by active pumps, in the
absence of any external potential difference $V=0$ and
concentration gradients $n^-=n^+$. Alternatively, it also applies
when the fluxes are produced by concentration gradients $n^- \neq
n^+$ in the absence of any external potential difference $V=0$,
provided that the two ions are different (we must require $G_1
\neq G_2$ or $D_1 \neq D_2$ so that $\sigma^\pm \neq 0$).

\noindent (ii) In addition to the renormalizations of the membrane
tension and bending modulus, a term proportional to $q^3$ in the
effective free energy is present. This term does not exist for an
equilibrium membrane in an electrolyte solution, as the
electrostatic potential only contains even powers of $q$. It is due
to the hydrodynamic couplings, which induce a non-analytic
dependence on $q$. Furthermore, the negative sign of $\Gamma$
suggests the possibility of an instability of the membrane
\cite{joanny}. To see this, let us first consider a free membrane.
There is no tension, and $\Sigma_t=\Sigma_0 + \Sigma=0$. The
membrane is unstable because the average squared undulation is
negative at low wave vector.  In the case of a vesicle, there is a
finite tension, which can be seen as a Lagrange multiplier ensuring
the conservation of its area. The vesicle is unstable if the total
tension is low enough to satisfy the condition
$\Sigma_t<\Gamma^2/4(K_0+K)$. This instability occurs at a wave
vector $q_c=-\Gamma / 2(K_0+K)$ suggesting a periodic deformation of
the membrane at this characteristic wave vector.  As expected, the
instability threshold for the membrane tension increases with the
ion fluxes or the applied potential difference.

\noindent (iii) The value of the renormalized membrane tension
$\Sigma$ can also be obtained from the zeroth order electrostatic
potential, with a simple mechanical argument.  Indeed, the surface
tension is related to the Maxwell stress by \cite{widom}:
\begin{equation}\label{surfacetension}
  \Sigma= -\epsilon \int_{-L/2}^{L/2} (E_z^{(0)})^2(z) dz + \frac{\epsilon L}{2} \left[
  (E_z^{(0)})^2(z \rightarrow \infty) + (E_z^{(0)})^2(z \rightarrow -\infty) \right].
\end{equation}
Note that in the r.h.s. of Eq.\ref{surfacetension}, the second
term cancels a contribution proportional to $L$ in the first term,
which originates from the existence of a pressure gradient in the
fluid.

\noindent (iv) For order-of-magnitude estimates, we obtain with
the parameters reported in Fig~\ref{fig:2}, $\Sigma = 3.2 \cdot
10^{-16}$ Jm$^{-2}$, $\Gamma = -10^{-24}$ Jm$^{-1}$ and $K =
10^{-13}k_BT$ . Although the ion flux is consequent and typical of
ion channels, the moduli $\Sigma$, $\Gamma$ and $K$ are very small
due to the strong dependance of these moduli on $\kappa^{-1}$,
which is here only 2.3nm. As we show below, these low values
also reflect the fact that sofar, we have neglected the bilayer character
of the membrane and its finite capacitance.

\section{Bilayer of finite thickness
and finite dielectric constant}

In this section, we extend the calculation to a bilayer of finite
thickness $d$ and dielectric constant $\epsilon_m<\epsilon$. There
is then an electrical coupling between the membrane and the
surrounding electrolyte, with a strength measured by the parameter
$t=\epsilon_m/(\kappa d \epsilon)$~\cite{helfrich,andelman}. For
equilibrium membranes, the importance of this coupling is
discussed in refs.~\cite{kleinert,chou}. For the sake of
simplicity, we only discuss the case of symmetric electrolytes:
$n^-=n^+$, $D_1=D_2=D$, and $G_1=G_2=G$ because we intend to
provide the full solution and more details in a longer paper. We
denote by $\phi_m$ the internal potential and $\phi$ the electrolyte
potential. When $t\neq0$, the boundary conditions at the membrane
are modified, they are now:
\begin{equation}\label{BC1}
\begin{array}{c}
  \epsilon \partial_z
\phi^{(0)}(z \to \pm d/2)=\epsilon_m \partial_z
\phi_m^{(0)}(z \to \pm d/2), \\
  \phi^{(0)}(z \to \pm d/2)=\phi_m^{(0)}(z \to \pm d/2),
\end{array}
\end{equation}
where the first equation is the continuity condition for the
normal electric displacement and the second equation is the
continuity condition of the potential. With these assumptions, we
find that the current-voltage relation given in Eq.~\ref{current}
still holds for any value of $t$. Furthermore, the electric field
for $z>d/2$ (resp. $z<-d/2$) is $E_z^{(0)}(z)=-\sigma /\epsilon
-\tilde{\sigma} /\epsilon \exp{ \left( \kappa \left( z+d/2 \right)
\right)}$ (resp. $E_z^{(0)}(z)=-\sigma /\epsilon -\tilde{\sigma}
/\epsilon \exp{ \left( \kappa \left( -z+d/2 \right) \right)}$),
while the internal field is constant and equal to $E_m=-(\sigma +
\tilde{\sigma})/(t \kappa d \epsilon)$. We have introduced
$\sigma$ the surface charge on the positive electrode, $\sigma$,which is
related to the electrical current $i=i_1=i_2$ in a way similar to
Eq.~\ref{surfacechargedef} by $\sigma=-2i/\kappa^2 D$, and
$\tilde{\sigma}$ defined by $\tilde{\sigma}=\int_0^\infty
\rho^+(z) dz$. These two surface charges are related by
$\tilde{\sigma}/\sigma=(D \kappa^3 \epsilon t - 2G)/2(2t+1) G$.
Note that only the diffusion time of the ions within a Debye layer
enters in $\sigma$ whereas $\tilde{\sigma}$ also contains the RC
characteristic time of the membrane.

Using Eq.~\ref{surfacetension} and the equations above,
the surface tension can now be written as the sum of an
internal contribution $\Sigma_{in}=-\epsilon_m d E_m^2=-(\sigma +
\tilde{\sigma})^2/t \kappa \epsilon$ and an external contribution
$\Sigma_{out}=\left(-\tilde{\sigma}^2-4 \sigma \tilde{\sigma} +
\kappa d \sigma^2 \right)/\epsilon \kappa$. The negative
contribution $\Sigma_{in}$ is known as the Lippmann
tension~\cite{sachs} and is usually larger in absolute value than
$\Sigma_{out}$. Because of $\Sigma_{in}$, the total membrane
tension $\Sigma_0+\Sigma_{in}+\Sigma_{out}$ can become negative at
some critical value of the internal field $E_m$, which leads to
the instabilities discussed in ref.~\cite{pierre}. As discussed in
the previous section, the surface tension $\Sigma$ can also be
recovered by solving Stokes equation at first order in the
membrane height with the appropriate electrostatic forces as
source terms. Care must be used here as the usual definition of
the (bulk) electrostatic force $\rho \vE$ holds only inside a
medium with a homogeneous dielectric constant, but not at the
boundaries between two dielectric media, where the jump in
dielectric constant is associated with a localized force. These
difficulties can be resolved by calculating the electrostatic
force directly with the Maxwell stress tensor \cite{landau}. From
the solution of Stokes equation, the velocity is obtained
everywhere in the domain $|z|>d/2$. By extrapolating this velocity
at $z=0$, an effective free energy of the same form as in
Eq.~\eqref{freeenergy} is obtained, with the same tension $\Sigma$
as calculated from Eq.~\ref{surfacetension}. We also obtain
the moduli $\Gamma = \sigma \tilde{\sigma}
\left( 8 + \kappa^2 d^2 + 4 \kappa d \right)/ 2 \kappa^2 \epsilon$
and $K$ which is a complicated quadratic function of $\sigma$ and
$\tilde{\sigma}$.

We now discuss the
limit of small conductance $G \rightarrow 0$ which is of experimental relevance.
In this limit,
$\sigma=0$ since there is no current in the medium $i=0$. The
membrane is a capacitor of surface charge
$\tilde{\sigma}=-\epsilon_m E_m=V \epsilon \kappa t / (1+2t)$.
This means that the equivalent circuit is made of three planar
capacitors in series, one being the membrane (of capacitance per
unit area $\epsilon_m/d$) and the other two corresponding to the
Debye layers on each side (of capacitance $\epsilon \kappa$ per
unit area), making up a total capacitance $C=\epsilon \kappa
t/(2t+1)$. Now $\Sigma_{in}=-\tilde{\sigma}^2/ t \kappa \epsilon$,
$\Sigma_{out}=-\tilde{\sigma}^2/ \kappa \epsilon$, and $\Gamma=0$.
In this limit, $\Gamma=0$ since the system is at
equilibrium and therefore the effective free energy must be of the
Helfrich form. With the values of the parameters used in
Fig.~\ref{fig:2} except for $G$ which we take to be G=$0$ and
$d=5$nm, $\epsilon_m/\epsilon = 1/40$, we obtain $t=3.3 \cdot
10^{-3}$ and $\kappa d=7.4$, $\Sigma_{in} = -8.4 \cdot 10^{-6}$
Jm$^{-2}$, $\Sigma_{out}=-1.0 \cdot 10^{-7}$ Jm$^{-2}$, $\Gamma =
0$ and $K = 0.011 k_B T$. If we use instead $G=10
\Omega^{-1}$/m$^2$ a typical value for ion channels, the order of
magnitude of the tension and bending modulus are unchanged and a
small value of $\Gamma=8.2 \cdot 10^{-20}$ Jm$^{-1}$ is found.
This indicates that the capacitor model with $G=0$ is a good
approximation to calculate the moduli in this case. The importance
of capacitive effects is confirmed by the fact that the values of
the moduli obtained here are much larger than the ones obtained
previously in the case of zero thickness. Furthermore, by varying
the ionic strength in the case where ion transport is present with
$G \neq 0$, we have found that the capacitor model holds at high
ionic strength but become invalid at low ionic strength where ion
transport has a stronger impact on the moduli. This will be discussed in a future work.

\section{Conclusions}
In conclusion, we have analyzed the steady-state fluctuations of a
membrane driven out-of-equilibrium by an applied DC electric field
or by concentration gradients. One of our most notable results is
the presence of a new term proportional to $q^3$ in the
fluctuation spectrum that we interpret as arising from
hydrodynamics couplings. For a free membrane or for a vesicle of
zero thickness and negligible dielectric constant, we have found a
negative value for $\Gamma$ which gives rise to a finite
wavelength instability. For a bilayer, we recover a known zero
wavevector instability, when the surface tension becomes negative.
Since $\Gamma$ is positive in this case, the other instability at
finite wavelength is absent for a bilayer. We have also analyzed
the role of capacitive effects, and found that they lead to a
dominant contribution in the electrostatic part of the surface
tension and bending modulus at high ionic strength. The analysis
presented here suggests directions for future study. In
particular, it would be interesting to go beyond the linear
approximation for the electrostatic potential and the ion
concentrations, and discuss with more details the implications for
the dynamics of active membranes containing for instance gated or
mechano-sensitive ion channels.

\acknowledgments We thank D.~Andelman for pointing out to us
ref.~\cite{duplantier}, P.~Sens, H.~Aranda-Espinoza,
J.B.~Fournier, A.~Ajdari, and J.~Prost for many useful
discussions. M.~C.~L. acknowledges financial support of a Human
Frontier Science Foundation grant.


\begin{thebibliography}{0}


\bibitem{Edidin03} E. Edidin, Nat.\ Rev.\ Mol. Cell. Biol.{\bf 4}, 414
  (2003).

\bibitem{pincus}P.\ Pincus, J.\ F.\ Joanny, D.\ Andelman, Europhysics letters {\bf 11},
763 (1990).

\bibitem{helfrich}
M.\ Winterhalter and W.\ Helfrich, J. Phys. Chem., {\bf 92}, 6865
(1988).

\bibitem{duplantier}
 R.\ E.\ Goldstein, A.\ I.\ Pesci, V.\ Romero-Rochin, Phys.\ Rev.\ A {\bf 41}, 5504
(1990).

\bibitem{girard} P.\ Girard, F.\ J$\ddot{u}$licher, J.\ Prost, Eur. Phys. J. E Soft Matter {\bf 14}, 387 (2004).

\bibitem{PRE_JBManneville}
J.-B.\ Manneville, P.\ Bassereau, S.\ Ramaswamy, and J.\ Prost,
Phys.\ Rev.\ E {\bf 64}, 021908 (2001).

\bibitem{leonetti} M.\ Leonetti, E.\ Dubois-Violette, and F.\ Hombl\'e,
PNAS {\bf 101}, 10243 (2004).

\bibitem{lacoste}
D.\ Lacoste and A.\ W.\ C.\ Lau, Europhys. Lett. , {\bf 70} (3),
418 (2005).

\bibitem{lomholt}
M.\ A.\ Lomholt, Phys. Rev. E, {\bf 73}, 061913 (2006).

\bibitem{armand}
A.\ Ajdari, Phys. Rev. Lett., {\bf 75}, 755 (1995).

\bibitem{hille} B.\ Hille ``Ion Channels of Excitable Membranes'', Sinauer
  Press, Sunderland MA (2001).

\bibitem{kandel} E.\ Kandel, J.\ Schwartz and T.\ Jessel, ``Principles of
neural science'', Mac Graw Hill, Ney York (2001).

\bibitem{seifert}
For an extensive review, see U.\ Seifert, Adv.\ Phys.\ {\bf 46},
13 (1997).

\bibitem{joanny}
S.\ T.\ Milner, J.-F.\ Joanny and P.\ Pincus, Europhys. Lett.,
{\bf 9}, 495 (1989).

\bibitem{widom}
J.\ S.\ Rowlinson and B. Widom, {\em Molecular Theory of
Capillarity} (Oxford, 1982).

\bibitem{andelman}
D.\ Andelman, {\em Electrostatic properties of membranes: the
Poisson-Boltzmann Theory in Handbook of Biological Physics edited
by R. Lipowsky and E. Sackmann} (1995).

\bibitem{kleinert}
M.\ Kiometzis and H.\ Kleinert, Phys. Lett. A, {\bf 140}, 520
(1989).

\bibitem{chou}
T.\ Chou, M.\ V.\ Jaric and E.\ Siggia, Biophysical Journal, {\bf
72}, 2042 (1997).

\bibitem{sachs}
P-C.\ Zhang, A.\ M.\ Keleshian, and F.\ Sachs, {\bf 413}, Nature,
428 (2001).

\bibitem{pierre}
P.\ Sens and H.\ Isambert, Phys. Rev. Lett., {\bf 88}, 128102
(2002).

\bibitem{landau}
L.\ D.\ Landau, E.\ M.\ Lifshitz and L.\ P.\ Pitaevskii, {\em
Electrodynamics of continuous media} (Elsevier, New York, 2002).


\end{thebibliography}
\end{document}